\documentclass[11pt]{article}
\pdfoutput=1
\usepackage{amsbsy,amsmath,amsthm,amssymb,graphicx,verbatim}
\usepackage{setspace, float, subfig}
\usepackage[longnamesfirst,sort]{natbib}

\usepackage{multirow}

\usepackage{geometry}
\theoremstyle{remark}

\newcommand{\sX}{\mathsf{X}}
\newfont{\msbm}{msbm10 at 11pt}
\newcommand {\R} {\mathbb{R}}
\newcommand {\Z} {\mbox{\msbm Z}}

\begin{document}
\onehalfspacing

\title{A practical sequential stopping rule for high-dimensional MCMC and its application to spatial-temporal Bayesian models}

\author{Lei Gong \\ Department of Statistics \\ University of California, Riverside \\ {\tt lei.gong@email.ucr.edu} \and James M. Flegal \\ Department of Statistics \\ University of California, Riverside \\ {\tt jflegal@ucr.edu} }

\date{\today}

\maketitle

\begin{abstract}
A current challenge for many Bayesian analyses is determining when to terminate high-dimensional Markov chain Monte Carlo simulations.  To this end, we propose using an automated sequential stopping procedure that terminates the simulation when the computational uncertainty is small relative to the posterior uncertainty.  Such a stopping rule has previously been shown to work well in settings with posteriors of moderate dimension.  In this paper, we illustrate its utility in high-dimensional simulations while overcoming some current computational issues.  Further, we investigate the relationship between the stopping rule and effective sample size.  As examples, we consider two complex Bayesian analyses on spatially and temporally correlated datasets. The first involves a dynamic space-time model on weather station data and the second a spatial variable selection model on fMRI brain imaging data. Our results show the sequential stopping rule is easy to implement, provides uncertainty estimates, and performs well in high-dimensional settings.

\end{abstract}

\section{Introduction}
\label{intro}

Markov chain Monte Carlo (MCMC) simulations are commonly employed in a Bayesian context to estimate features of a posterior distribution by constructing a Markov chain with the target as its stationary distribution. A fundamental challenge is determining when to terminate the simulation, especially for the often high-dimensional problems encountered in modern MCMC.  For instance, the visual inspection of trace plots and running means \citep[see e.g.][]{fleg:jone:2011} is extremely challenging in high-dimensions.  Further, convergence diagnostics \citep[see e.g.][]{cowl:carl:1996} were designed for problems of at most moderate dimension and can be essentially impossible to implement in high-dimensions.  Given these problems, most practitioners resort to using a fixed-time rule to terminate the simulation.  That is, the procedure terminates after $n$ iterations where $n$ is determined heuristically.  In this paper, we present a simple sequential stopping rule applicable for high-dimensional MCMC.

As application, we consider the analysis of large spatially and temporally datasets routinely collected by the scientific community.  An unprecedented framework to effectively incorporate spatial-temporal associations is to build dependencies in different stages of Bayesian hierarchical models \citep{bane:carl:gelf:2004}, which often involves the implementation of high-dimensional MCMC. There is considerable literature in this direction, for example, \cite{huer:sans:stro:2004} develop a time-varying regression model for studying ozone levels; \cite{gelf:bane:game:2005} propose spatial process modeling for dynamic data with an application to climate data; \cite{finl:bane:gelf:2012} use Gaussian predictive processes to model large space-time data; \cite{wool:jenk:brad:smit:2004} implement a fully spatio-temporal model for the noise process in fMRI data; \cite{smit:fahr:2007} and \cite{lee:jone:caff:bass:2011} develop spatial Bayesian variable selection models to brain image study.

With important economic, ecological and public health implications, these analyses require accurate assessment of their inferential uncertainties. However, few of these studies, which often involve thousands of parameters, carefully describe the stopping criterion utilized.  Among them, some use convergence diagnostics \citep[see e.g.][]{gelf:bane:game:2005} and some report Monte Carlo standard errors (MCSEs) to assess the quality of estimates \citep[see e.g.][]{lee:jone:caff:bass:2011}.  We assume the rest employ a fixed-time stopping rule where $n$ is determined heuristically.  Unfortunately, choosing too small an $n$ can lead to inaccurate statistical inference.

As mentioned, many practitioners utilize convergence diagnostics and visual inspections to evaluate if the chain has been run long enough.  Unfortunately, these methods are barely tenable in truly high-dimensional settings.  For example, as stated in \cite{goss:auer:fhar:2001}, ``With this high-dimensional data, convergence diagnostics were reduced to a selection of randomly chosen parameter chains''.

Instead, we advocate terminating the simulation using a fixed-width stopping rule (FWSR), which are easy to implement and theoretically justified.  The main idea is to terminate the simulation when an estimate is sufficiently accurate for the scientific purpose of the analysis.  That is, the simulation is terminated the first time a confidence interval width for a desired quantity is sufficiently small.  Hence, the the total simulation effort will be random with these procedures. 

A FWSR, first studied in MCMC settings by \cite{jone:hara:caff:neat:2006}, stops the simulation when the width of a confidence interval is less than a pre-specified value $\epsilon$.  Further, \cite{jone:hara:caff:neat:2006} and \cite{fleg:hara:jone:2008} show FWSR is superior to using convergence diagnostics as stopping criteria.  \cite{fleg:gong:2013} propose variants known as relative FWSRs that eliminate the need to specify an absolute value $\epsilon$. Further, relative FWSRs are more practical in multivariate settings since a single $\epsilon$ can be used for multiple parameters without apriori knowledge.  In particular, they advocate the use of relative standard deviation FWSR for Bayesian computations.  In short, this rule terminates the simulation when the computational uncertainty of an estimate is small relative to its posterior uncertainty.

In this article, we further investigate the finite sample properties of the relative standard deviation FWSR and propose necessary modifications for applications in high-dimensional settings. The proposed modifications are driven by computational issues providing significant improvement with minor tradeoffs.  As we see later, the benefits of our modifications are a significant reduction in the required computer memory and improved computational efficiency. To our best knowledge, there are no previous attempts to formally address how long to run a MCMC simulation in such high-dimensional settings. Specifically, we extend the previous application \citep{fleg:gong:2013} of the stopping rule for estimating tens of parameters to a spatial Bayesian dynamic model with hundreds of parameters and a more complicated Bayesian fMRI model with thousands of parameters.  Further, we compare our results to a convergence diagnostic used as a stopping criterion and show the latter tends to terminate the simulations prematurely. We also establish a connection between the relative standard deviation FWSR and an alternative effective sample size (ESS) calculation.

The two distinct high-dimensional Bayesian hierarchical analyses considered here are (i) the univariate dynamic space-time regression models introduced by \cite{gelf:bane:game:2005} applied to weather station data collected over the northeastern United State between 2000 and 2005 \citep{finl:bane:gelf:2012} and (ii) the spatial variable selection models proposed by \cite{lee:jone:caff:bass:2011} applied to the StarPlus fMRI datasets \citep{carp:just:kell:eddy:thul:1999, kell:just:sten:2001, wang:mitc:2002}. Both applications clearly demonstrate the potential of the relative standard deviation FWSR in general high-dimensional settings. Moreover, they illustrate the rule is easily implemented in an almost automated fashion while providing uncertainty estimates with confidence. 

The rest of the paper is organized as follows. Section~\ref{sec:stop} formally introduces the relative standard deviation FWSR, proposes modifications for modern applications, and illustrates connections with an ESS calculation. Section~\ref{sec:appl} details application to two high-dimensional MCMC inferences for Bayesian hierarchical models, where the model, experimental dataset and several comparative studies are summarized. Section~\ref{sec:disc} concludes with a discussion.

\section{Sequential stopping procedure} \label{sec:stop}

Suppose we want to make inference about a probability distribution $\pi$ with support $\sX \subseteq \mathbb{R}^{d}$, $d \ge 1$. In general, we denote $\pmb{\theta} = (\theta_1, \dots, \theta_p)^{T} \in \R^{p}$, $p \ge 1$ as a target parameter of interest with respect to $\pi$. Note that $p$ can be smaller or larger than $d$, with large values of either indicating a high-dimensional setting. In this paper, we will restrict our attention to 
\begin{equation*}
\pmb{\theta} := E_{\pi} [ g(X) ] = \int_{\sX} g(x) \pi(dx) \; , 
\end{equation*}
where $g: \sX \to \R^{p}$. 

Unfortunately, in most practical settings we cannot calculate $\pmb\theta$ analytically and frequently $\pi$ is such that MCMC is the only viable technique for estimating $\pmb\theta$.  The basic MCMC methods entail constructing a time-homogeneous Harris ergodic Markov chain $X = \left\{  X^{(0)} ,  X^{(1)} , \ldots\right\}$ on state space $\sX$ with $\sigma$-algebra $\mathcal{B} = \mathcal{B}(\mathsf{X})$ and invariant distribution $\pi$ \citep{robe:case:2004}.

Suppose we simulate $X$ for $n$ iterations, where $n$ is finite. Let 
\[
\pmb{Z} (n) := \frac{1}{n} \sum_{i=0}^{n-1} g\left( X^{(i)} \right)= (Z_{1} (n) , \dots, Z_{p} (n) )^{T}
\]
be an estimator of $\pmb\theta$ from the observed chain. Under certain regularity conditions \citep{jone:2004, chan:geye:1994, robe:rose:2004, tier:1994}, we can obtain a marginal Markov chain central limit theorem (CLT) for the sampling distribution of an unknown Monte Carlo standard error, $\pmb{Z} (n) - \pmb{\theta}$,
\begin{equation}
\label{eq:clt}
\sqrt{n} \left( Z_{i} (n) - \theta_i \right) \stackrel{d}{\to} \; \text{N} \left( 0, \sigma_{i} ^2 \right)
\end{equation}
as $n \to \infty$ where $\sigma_{i} ^2 \in (0, \infty)$, $i = 1, \dots, p$. One could also consider a multivariate Markov chain CLT for $\pmb{Z} (n) - \pmb{\theta}$. However, the often high-dimensionality of the associated asymptotic covariance matrix creates additional challenges and extracting useful information from it is a direction of future research.

For $i = 1, \dots, p$, let $\hat{\sigma}_{i}(n)$ denote an estimator of $\sigma_{i}$.  Then the CLT allows construction of a $(1-\delta)100\%$ marginal confidence interval for $\theta_i$, $i = 1, \dots, p$, with width
\begin{equation}
\label{eq:hw}
w_{i} (n,\delta) = 2 z_{\delta / 2} \frac{\hat{\sigma}_{i} (n)}{\sqrt{n}} 
\end{equation}
where $z_{\delta / 2}$ is a critical value from the standard Normal distribution.  The width at \eqref{eq:hw} allows analysts to report the uncertainty in their estimates and users to assess the practical reliability. Moreover, we can use it to construct sequential FWSRs that terminates the simulation when the $w_{i}$s fall below a specific value.

\subsection{A relative fixed-width stopping rule}

Suppose $\epsilon$ is a pre-specified value, then the simplest FWSR terminates the simulation when $w_{i} < \epsilon$ for all $i = 1, \dots, p$.  Asymptotic validity of such a rule was established by \cite{glyn:whit:1992} and first used in MCMC simulations by \cite{jone:hara:caff:neat:2006}.  Asymptotic validity is important because it ensures the simulation will terminate w.p.1 and the resulting confidence interval will have the right coverage probability (as $\epsilon \rightarrow 0$).  Unfortunately, such a rule is difficult to implement in high-dimensional settings without apriori knowledge of the magnitudes of the components in $\pmb \theta$.  Further, a single $\epsilon$ value is unlikely to be suitable across multiple dimensions.

Instead, we advocate the use of what is known as a relative standard deviation FWSR proposed by \cite{fleg:gong:2013}. The main idea is to terminate the simulation when an estimators computational uncertainty is small relative to its posterior uncertainty.  As we will illustrate, this is equivalent to terminating the simulation when the ESS is sufficiently large.

First, we require a bit more notation.  Let $\lambda_{i}^2$ denote the posterior variance associated with $\theta_{i}$.  That is, if an i.i.d.\ sample from $\pi$ were available, then $\lambda_{i}^2$ is the asymptotic variance in the CLT associated with $\theta_{i}$.  It is important to note in general that $\sigma_{i} ^2 \neq \lambda_{i}^2$ due to correlation in the Markov chain.  For estimation of $E_{\pi} [ g(X) ] $, it is easy to verify that $\lambda_{i}^2$ is the $i$-th diagonal element of $Var_{\pi} [ g(X) ]$.  We further suppose $\hat{\lambda}_{i}^2 (n)$ is an estimator of $\lambda_{i}^2$, usually 
\[
(\hat{\lambda}_{1}^2 (n), \dots, \hat{\lambda}_{p}^2 (n) )^{T} = \frac{1}{n-1} \sum_{i=0}^{n-1} \left( g\left( X^{(i)} \right) - \pmb{Z} (n) \right)^2 \; .
\]
Note that exponentiation on a vector is taken element-wise.

A relative standard deviation FWSR terminates the simulation when the length of all the confidence intervals are less than an $\epsilon$th fraction of the magnitude of their posterior standard deviations.  That is, when $w_{i} < \epsilon \hat{\lambda}_{i} $ for all $i = 1, \dots, p$.  Formally, the time at which the simulation terminates is defined by
\[
T (\epsilon) = \sup_{i \in \{ 1, \dots, p \}} \inf \left\{ n \ge 0 : w_{i} (n,\delta) + p(n) \le \epsilon \hat{\lambda}_{i} (n) \right\} \; ,
\]
where $p(n) \ge 0$.  The role of $p(n)$ is to ensure that the simulation is not terminated prematurely based on a poor estimate of $\sigma_{i} ^2$.  A reasonable default is $p(n) = \epsilon I(n \le n^{*}) + n^{-1}$ \citep{glyn:whit:1992, jone:hara:caff:neat:2006}, where $n^{*}$ is the desired minimum simulation effort. The user-specified starting value $n^{*}$ is often based on the complexity of the problem at hand and $\epsilon$ reflects the desired accuracy for the analytical purpose.

Sufficient conditions for asymptotic validity of the relative standard deviation stopping rule are established in \cite{fleg:gong:2013}.  In short, we require three primary assumptions for $i = 1, \dots, p$.  First, the limiting process must satisfy a functional central limit theorem (FCLT).  
Second, the estimator of the associated asymptotic variance must be strongly consistent, that is $\hat{\sigma}^{2}_{i} (n) \to \sigma^{2}_{i}$ w.p.1 as $n \to \infty$. Finally, the estimator of the posterior variance must be strongly consistent, that is $\hat{\lambda}_{i}^2 (n) \to \lambda_{i}^2 > 0$ w.p.1 as $n \to \infty$.  While not trivial, one can establish these conditions in many complex practical MCMC settings.  An interested reader is directed to \cite{fleg:gong:2013}, \cite{fleg:jone:2010}, and \cite{jone:hara:caff:neat:2006}.  

The relative standard deviation FWSR is appealing to Bayesian practitioners because it provides a simple, yet informative automated stopping criterion applicable in multivariate settings.  First, one only needs to specify a relative $\epsilon$ and hence no prior knowledge about the magnitude of the parameters is needed. Second, a single $\epsilon$ will suffice in multivariate settings, whereas the other fixed-width approaches require a vector of values.  For small $p$, \cite{fleg:gong:2013} recommend using $\epsilon = 0.02$, which provides excellent results in a wide variety of examples.  However, one may adjust $\epsilon$ to reach a balance between accuracy and computational cost.

Once the simulation effort exceeds the starting value $n^{*}$, the frequency with which the criterion should be checked is still an open question.  Checking too often may substantially increase the computational burden. Instead, it is sufficient to check every $m$ iterations, where $m$ is a pre-specified gap determined by an estimated simulation effort.



\subsection{Variance estimation modification}

The MCMC community has expended considerable effort in establishing strongly consistent estimators for the asymptotic variance at \eqref{eq:clt} including batch means \citep{fleg:jone:2010, jone:hara:caff:neat:2006}, spectral variance estimation \citep{fleg:jone:2010} and regenerative simulation \citep{hobe:jone:pres:rose:2002, mykl:tier:yu:1995}. We consider only non-overlapping batch means (BM) since it is easy to implement and is feasible in high-dimensional settings.

In standard BM the output is broken into $a_n$ batches of equal size $b_n$.  Suppose the algorithm is run for a total of $n=a_n b_n$ iterations and define for $j=1,\ldots,a_n$
\begin{equation*}
\pmb{Y}_{j} = \frac{1}{b_n} \sum_{i=(j-1)b_n}^{jb_n-1} g\left( X^{(i)} \right) \; .
\end{equation*}
The BM estimate of the asymptotic variance from the CLT at \eqref{eq:clt} is
\begin{equation*}
\left( \hat{\sigma}_{1}^2(n), \dots, \hat{\sigma}_{p}^2(n) \right)^T = \frac{b_n}{a_n-1} \sum_{j=1}^{a_n} \left( \pmb{Y}_{j} - \pmb{Z} (n) \right)^{2}  \; .
\end{equation*}

\citet{jone:hara:caff:neat:2006} establish necessary conditions for $\hat{\sigma}_{i}^2(n) \to \sigma^{2}_{i}$ w.p.1 as $n \to \infty$ if the batch size and the number of batches are allowed to increase as the overall length of the simulation increases.  Setting $b_{n} = \lfloor n^{\tau} \rfloor$ and $a_{n} = \lfloor n / b_{n} \rfloor$, the regularity conditions require that $X$ be geometrically ergodic, $E_{\pi} |g|^{2+\epsilon_{1}+\epsilon_{2}} < \infty$ for some $\epsilon_{1} >0$, $\epsilon_{2} >0$ and $(1+\epsilon_{1}/2)^{-1} < \tau < 1$.  A common choice of $\tau=1/2$ has been shown to work well in applications \citep{jone:hara:caff:neat:2006, fleg:hara:jone:2008}. We denote the BM estimate with such sampling plan as the usual batch means (uBM) estimate.

Unfortunately, most sampling plans such as uBM require storage of the entire Markov chain to allow recalculations as the batch size $b_n$ grows with $n$.  Given a target vector of dimension $p$, this means a matrix of size $p \times n$ will have to be stored in the memory.  Clearly, computer memory soon becomes a serious issue, which one can solve by writing parts of the chain out of memory.  However, given the frequency that $T (\epsilon)$ is checked and the already computationally intense task of updating the chain, we prefer to simplify the problem.

To this end, we suggest a new sampling plan utilizing less memory while still providing a strongly consistent BM variance estimator.  Specifically, one could set $\tilde{b}_n = \sup \left\{ 2^k : 2^k \le n^{\tau}, k \in \Z^{+} \right\}$ and $\tilde{a}_{n} = \lfloor n / \tilde{b}_{n} \rfloor$. Notice $\tilde{b}_n$ is bounded by $n^{\tau}/2 \le \tilde{b}_n \le n^{\tau}$.  Hence, it is easy to establish strong consistency for $\hat{\sigma}_{i}^2(n)$ with such a sampling plan using results in \cite{jone:hara:caff:neat:2006} and \cite{bedn:latu:2007}.  We denote this BM estimate with $\tilde{b}_n$ as the alternative batch means (aBM) estimate.

Notice that $\tilde{b}_n$ increases by doubling the batch size only, i.e.\ in the form of $\left\{ 2, 4, 8, \dots, 2^{k}, \dots \right\}$. It then becomes possible to record only the $\pmb{Y}_{j}$s and merge every two batches by averaging their means when the batch size increases twofold. The size of the required storage then reduces significantly from $O(n)$ to $O(\tilde{a}_{n}) = O(n^{1 - \tau})$.  Moreover, at each $T (\epsilon)$ checking point, the calculation time decreases since the batches are already in memory. In practice, this change saves a noteworthy amount of computational effort as we will illustrate later.  Finally, using the new sampling plan with $T (\epsilon)$ requires a standard recursive recalculation of $\hat{\lambda}_i(n)$ as $n$ increases.  An interested reader is directed to the technique studied by \cite{bies:1977}.  

In the proposed sampling plan, we are only storing the current state of the chain and the batch means.  Then in some sense, the unit of interest is per batch rather than per iteration. Thus, a natural adjustment to the frequency with which $T (\epsilon)$ should be checked is to examine the criterion every $m$ batches instead of every $m$ iterations.  As before, $m$ is pre-specified by the user but is likely much smaller than used previously.  The gap between checks is then $m$ batches, or equivalently $m \tilde{b}_n$ iterations.  Hence, the iterations between checks automatically increases in accordance to the magnitude of the number of iterations without having to estimate the total simulation effort beforehand. In this regard, a similar increase between checks also performs well in general sampling plans. 

The two proposed modifications fit naturally with each other and enable implementation of relative standard deviation FWSR in high-dimensional settings. In addition, they yield improvements in memory requirements, computational efficiency measured by clock time, and stopping procedure automation. One drawback of the new sampling plan is not applicable to estimation problems that require storing the entire chain, hence it is best served in expectation estimation.

\subsection{Connections with effective sample size}

Given $n$ iterations in a Markov chain, the ESS measures the size of an i.i.d.\ sample with the same standard error.  One way to define ESS is described in \cite{kass:1998} and \cite{robe:case:2004}, for $i = 1, \cdots, p$,
\[
\text{ESS}_i = \frac{n}{\tau_i} = \frac{n}{1+2 \sum_{k=1}^{\infty} \rho_{k}(\theta_i)},
\]
where $\rho_{k}(\theta_i)$ is the autocorrelation of lag $k$ for $\theta_i$. This calculation is implemented in many R packages, such as {\tt coda} \citep{best:cowl:vine:1995} and {\tt mcmcse} \citep{fleg:hugh:2012}. 

An alternative approach to define ESS is through the ratio of $\sigma_{i} ^2$ and $\lambda_{i}^2$.  Given the previous strongly consistent estimates, we have as $n \to \infty$
\begin{equation}
\label{eq:ess}
\frac{n \hat{\lambda}_{i}^2 (n)}{\hat{\sigma}_{i}(n)} \to \frac{n \lambda_{i}^2}{\sigma_{i} ^2} = \text{ESS}_i.
\end{equation}
The two ESS calculations produce comparable results in various simulation studies with relatively small correlations. However, our ESS calculation is systematically smaller due to a larger variance estimate when sample is highly correlated.  Moreover, the alternative ESS calculation provides additional insights into the relative standard deviation FWSR. From the definition of $T (\epsilon)$, one can easily show that at termination
\begin{equation}
\label{eq:term}
\epsilon \hat{\lambda}_{i}^2 (n) \geq 2 z_{\delta / 2} \hat{\sigma}_{i}(n) / \sqrt{n} + p(n) \geq 2 z_{\delta / 2} \hat{\sigma}_{i}(n) / \sqrt{n}.
\end{equation}
Combining \eqref{eq:ess} and \eqref{eq:term}, we have
\begin{equation*}
\text{ESS}_i \geq \frac{4 z_{\delta / 2}^2}{\epsilon^2}, \; \forall i = 1, \cdots, p.
\end{equation*}
Hence, the relative standard deviation FWSR is equivalent to terminating a simulation when the smallest ESS is above pre-specified level.  For instance, setting $\epsilon = 0.05$ and $\delta = 0.05$ produces an estimated ESS at least $6147$.  That is, such a simulation would provide equivalent inference as $6147$ i.i.d.\ samples.

\subsection{An alternative stopping criterion}

Convergence diagnostics \citep[for a review see][]{cowl:carl:1996} are widely employed by practitioners as a stopping criteria. Particularly, we are interested in the Geweke diagnostic (GD)  from \cite{gewe:1992}, which we will compare with relative standard deviation FWSR in the next section.  Our simulations use the GD implementation from the R package {\tt coda} \citep{best:cowl:vine:1995}. The GD is based on a hypothesis test that the mean estimates of two non-overlapping parts of the Markov chain have converged.  As a rule of thumb, \cite{gewe:1992} suggested to take first 0.1 and last 0.5 proportions of the Markov chain. The resulting test statistic is univariate by its nature and the $z$-score is constructed as follows,
\[
Z = \frac{\bar{x}_1 - \bar{x}_2}{\sqrt{ \hat{s}_1(0)/n_1 + \hat{s}_2(0)/n_2 }},
\]
where $\bar{x}_1, \bar{x}_2$ are the sample average and $\hat{s}_1(0), \hat{s}_2(0)$ are the consistent spectral density estimates at zero frequency for the two parts of the Markov chain, respectively.

The GD requires a single Markov chain, which is close in spirit to the current work.  It is also more practical in high-dimensional settings than the popular Gelman-Rubin diagnostic \citep{gelm:rubi:1992a, broo:gelm:1998}, which requires parallel chains. \cite{jone:hara:caff:neat:2006} note that the GD is based on a Markov chain CLT and hence does not apply more generally than a FWSR that is based on the calculation of MCSE.

\section{Applications} \label{sec:appl}

In this section, we evaluate the performance of the relative standard deviation FWSR and the proposed modifications using modern Bayesian applications. Particularly, we look at the spatial Bayesian dynamic models \citep{gelf:bane:game:2005} with application to the weather station data \citep{finl:bane:gelf:2012} and the spatial Bayesian variable selection models \citep{lee:jone:caff:bass:2011} on an experiment fMRI dataset \citep{carp:just:kell:eddy:thul:1999}. The modified stopping rule is implemented to terminate the MCMC simulations for posterior inference and comparative studies are given to illustrate its several advantages.

\subsection{Bayesian dynamic space-time model}

This application considers the monthly temperature data collected over 356 weather stations in the northeastern Unite States starting in January 2000 to September 2011, which is available in the R package {\tt spBayes} \citep{finl:bana:2013}. We fit the univariate Bayesian dynamic space-time regression model proposed by \cite{gelf:bane:game:2005} to a subset of the dataset for illustrative purposes. Specially, we only interested in the data recorded from 10 nearby weather stations in the year 2000. Note that the modeling approach is limited to the settings where space is continuous but time is taken to be discrete.

The response $y_t(s)$ denotes the temperature at location $s$ and time $t$. It is modeled through a measurement equation that provides a regression specification with a space-time varying intercept and serially and spatially uncorrelated zero-centered Gaussian disturbances as measurement error $\epsilon_t(s)$. Next a transition equation introduces a $p \times 1$ coefficient vector ${\pmb \beta}_t$, which is a strictly temporal component, and a spatial-temporal component $u_t (s)$. The overall model is given by
\[ y_t(s) = {\pmb x}_t (s)^{T} {\pmb \beta}_t + u_t (s) + \epsilon_t (s), \ t = 1, 2, \dots, N_t, \]
\[ \epsilon_t \sim N(0, \tau_t^2), \]
\[ {\pmb \beta}_t = {\pmb \beta}_{t-1} + {\pmb \eta}_t; \ {\pmb \eta}_t \sim N_p (0, \Sigma_{\eta}), \]
\[ u_t(s) = u_{t-1}(s) + w_t(s); \ w_t(s) \sim GP (0, C_t(\cdot, \psi_t)). \]
The $GP (0, C_t(\cdot, \psi_t))$ denotes a spatial Gaussian process with covariance function $C_t(\cdot; \psi_t)$. We specify $C(s_1, s_2; \psi_t) = \sigma_t^2 \rho (s_1, s_2; \phi_t)$, where $\psi_t = \{ \sigma_t^2, \phi_t \}$ and $\rho (\cdot; \phi)$ is an exponential correlation function with $\phi$ controlling the correlation decay and $\sigma_t^2$ represents the spatial variance component.

The prior specifications and MCMC schemes follow the {\tt spDynLM} function in the {\tt spBayes} package and we use it to sample from the Markov chain. Interested readers are directed to \cite{finl:bana:2013} for details. Particularly, we are interested in estimating the posterior mean of ${\pmb \theta} = \{ \tau_t^2, \sigma_t^2, \phi_t, {\pmb \beta}_t, \Sigma_{\eta}, u_t(s) \}$. The subset of data upon which we fit the model produces the parameters of interest ${\pmb \theta} $ where $p = 186$. 

We terminated simulations under several settings and conducted comparative studies of the outputs. Specifically, the relative standard deviation FWSR is implemented using two batch mean estimates (uBM and aBM) and two $\epsilon$ values (0.1 and 0.05).  Since the proposed sampling plan is applied, which leads to the batch size of the form $2^k$, $k \in \Z^{+}$, we set $n^{*} = 2^{14} = 16,384$ and added $20$ or $21$ batches between checks. The variants of the added batches is to ensure an even number of batches when checking the modified stopping rule. The nominal level of the coverage probabilities was set to 0.95. We compare our results to the GD with a threshold $p$-value 0.05 checked after 15,000 iterations in a effort to confirm the chain convergence.

\begin{table}[ht]
\begin{center}
\begin{tabular}{l|cccc}
Methods & Threshold & Length & ESS & CPU time\\
\hline
	uBM & $\epsilon = 0.1$ & $4.74E5$ & $2.22E3$ &  $4.43E3$ \\
	aBM & $\epsilon = 0.1$ & $4.06E5$ & $2.20E3$ & $2.37E3$ \\
	uBM & $\epsilon = 0.05$ & $2.49E6$ & $9.15E3$ & $1.04E4$ \\
	aBM & $\epsilon = 0.05$ & $2.24E6$ & $9.05E3$ & $4.39E3$ \\
	GD & p.v. = 0.05 & $1.50E4$ & $2.41E2$ &  $7.06E1$ \\
\hline
\end{tabular}
\caption{Summary statistics for uBM, aBM and GD. The ESSs are reported using its median. The CPU times are recorded in seconds.}
\label{tab:spB}
\end{center}
\end{table} 

\begin{figure}[thb]
\centering
\includegraphics[width=0.7\textwidth]{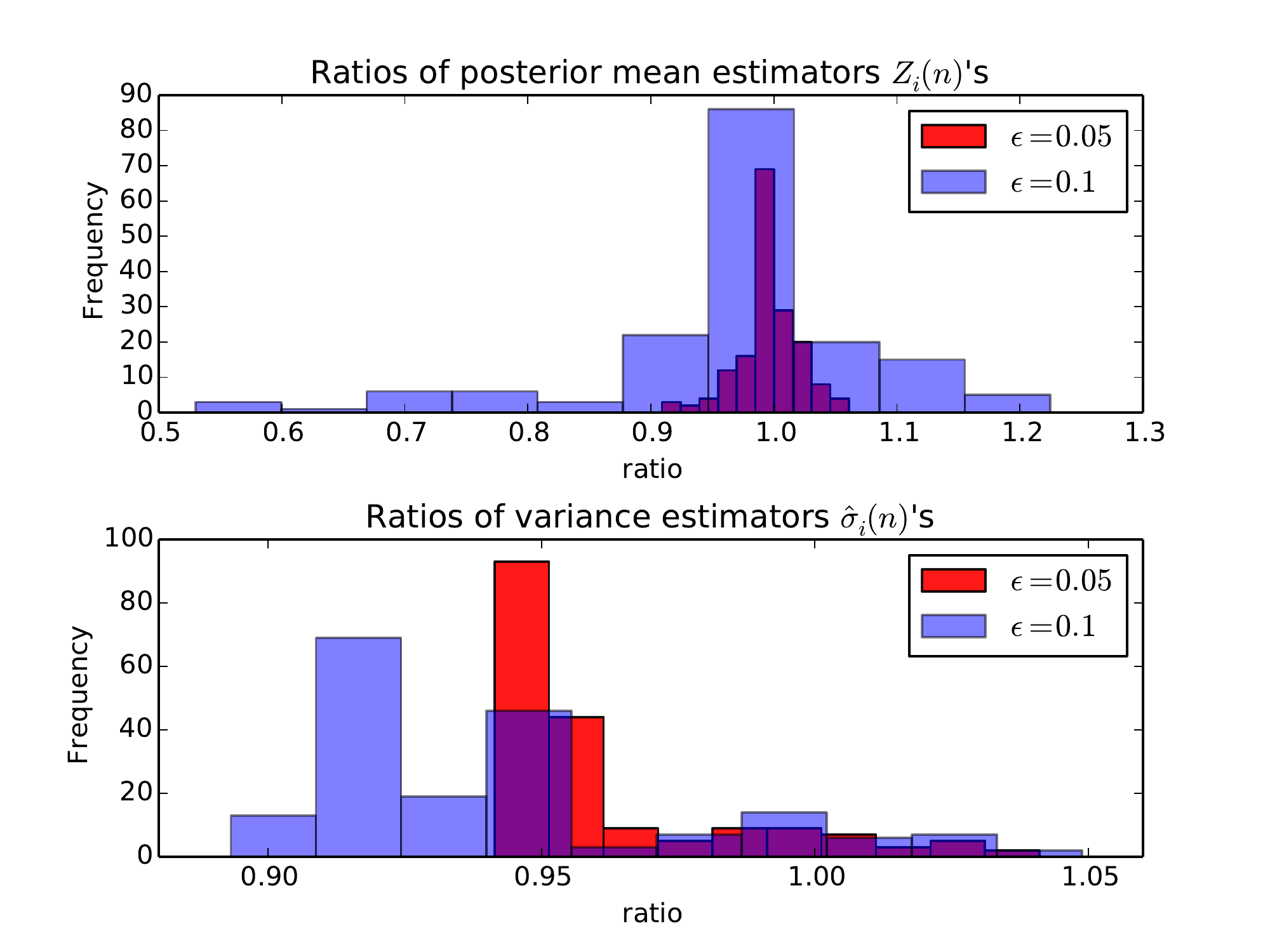}
\caption{The ratios of estimates obtained from aBM over those from uBM for $\epsilon = \{0.1, 0.05 \}$.}
\label{fig:ratio}
\end{figure}

Table~\ref{tab:spB} summarizes the comparative statistics for the stopping criteria. Note that ESS estimates a minimum equivalent sample size for $i = 1, \dots, p$ if an i.i.d.\ sampler was available. Among the results from the FWSR, we see that uBM and aBM perform similarly given $\epsilon$ in terms of the total simulation effort and ESS. However, notice that, aBM reduces the CPU time significantly as the total simulation effort increases. We further compare the 186 estimates obtained from uBM and aBM in terms of the ratios of estimated posterior means and variances under the two $\epsilon$ settings. Figure~\ref{fig:ratio} illustrates that there are no significant differences in the estimates, although one tradeoff is that the estimates $\hat{\sigma}_i(n)$s from aBM are biased downward slightly.  Given the advantages of aBM when $p$ is large where the uBM may not work, we advocate its use for relative standard deviation FWSR in high-dimensional settings. In the rest of the paper, we set aBM as the default sampling plan for $T(\epsilon)$ if not otherwise specified. 

\begin{figure}[thb]
\centering
\includegraphics[width=0.7\textwidth]{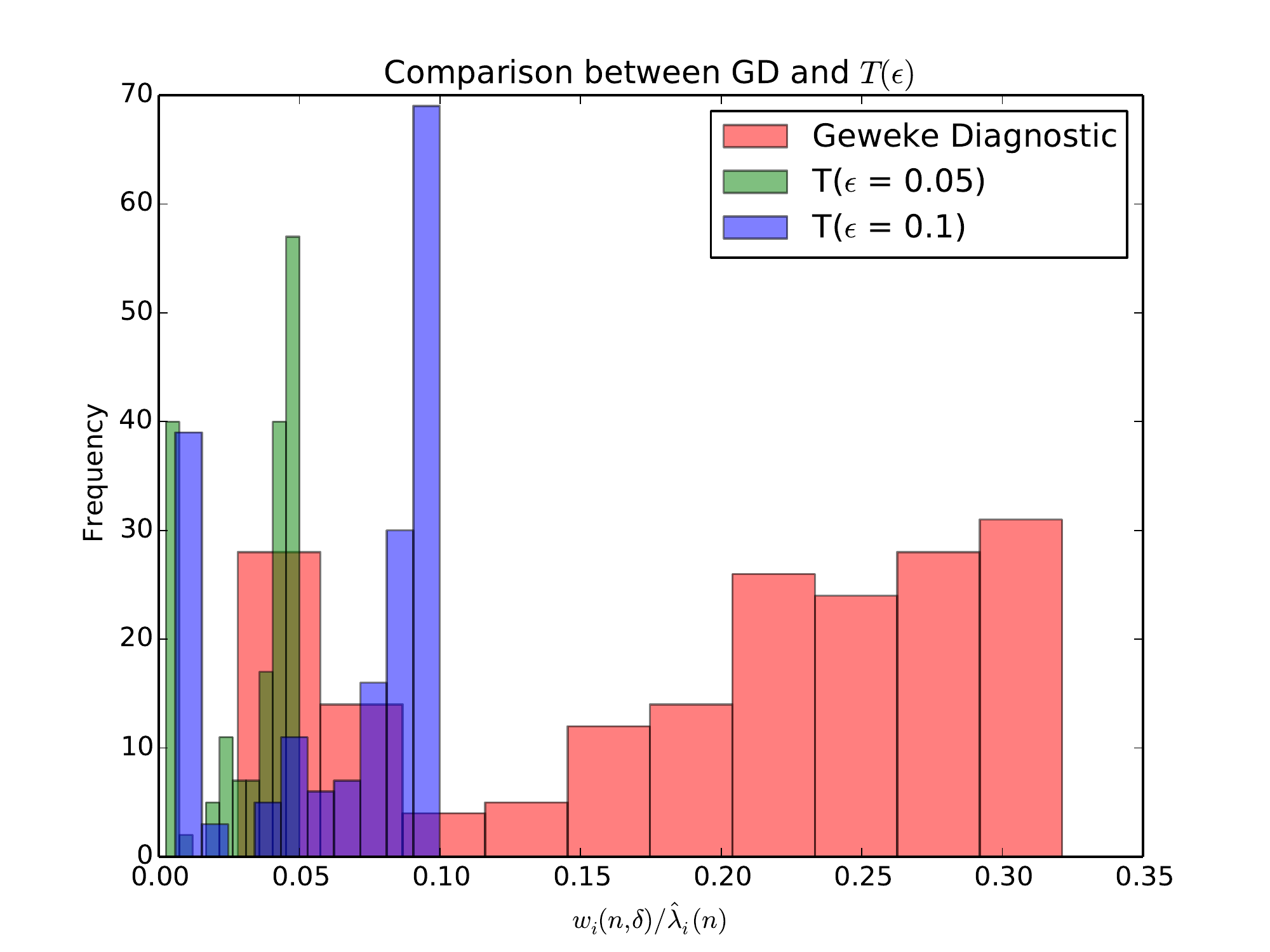}
\caption{The visualization of the comparison between GD and $T(\epsilon)$ in terms of the ratios $w_{i} (n,\delta) / \hat{\lambda}_{i} (n)$.}
\label{fig:gdratio}
\end{figure}

On the other hand, the total chain length and the ESS makes it clear that GD terminates the simulation too soon. Further, one can compare GD with $T(\epsilon)$ using the ratios $w_{i} (n,\delta) / \hat{\lambda}_{i} (n)$. Recall that $w_{i} (n,\delta)$s are the width of the resulting confidence intervals. These ratios serve as a measurement of the quality of our estimates. It assesses the computational uncertainty by using MCMC relative to the posterior standard deviation. Figure~\ref{fig:gdratio} visualizes the comparison of these ratios. It is clear the ratios from the FWSR $T(\epsilon)$ are significantly smaller and more concentrated than the ones from GD. Notice the cutoffs at 0.05 and 0.1 in Figure~\ref{fig:gdratio} are determined by $\epsilon$ values. These findings agree with \cite{cowl:carl:1996} that GD tends to be premature in diagnosing convergence.

\subsection{Spatial Bayesian variable selection model}

This application considers the Bayesian analysis of functional Magnetic Resonance Imaging (fMRI) study. It studies the physiological changes that accompany brain activation via the blood oxygenation level dependent (BOLD) signal contrast. During the course of a typical fMRI experiment, a single patient performs a set of tasks in response to one or several external stimulus while a series of three dimensional brain images are acquired. Our goal is to detect activated brain regions associated with external stimulus through the image intensities. Imagine that the patient's brain can be divided into tiny voxels on a 3D regular lattice. The time series BOLD response is collected at each voxel resulting in enormous observations of spatio-temporally correlated structures. The analysis fMRI data often involves ultrahigh-dimensional Bayesian models and extensive computation. 

Particularly, we are interested in the StarPlus experiment introduced by \cite{carp:just:kell:eddy:thul:1999}. The experiment was designed to investigate brain activities related to high level cognition, i.e. language comprehension and visuospatial processing. It involved eight subjects (four males, four female) with multiple trails. Specifically, we restrict our attention to subject 04847 and trail No.\ 3. In this trail (approximately twenty-seven seconds), the subject was shown a picture such as '$+ \over *$' for four seconds. Then the picture was replaced by a blank screen for another four seconds. Later, the second stimulus, i.e. a sentence such as "the star sign is above the plus sign", was presented for four seconds, or until the subject pressed a button indicating if the sentence correctly described the picture, whichever came first. A rest period of fifteen seconds was added after the sentence was removed. Snapshots were taken every $0.5$ seconds resulting in about $54$ images. Data were preprocessed using standard techniques such as slice timing and spatial smoothing \citep[for a review see][]{lind:2008} and were registered in standardized space with $64 \times 64 \times 8$ dimensions for $54$ time points. Snapshot at each time point has a dimensionality 4,698 voxels. 

\begin{figure}[thb]
\centering
\includegraphics[width=0.7\textwidth]{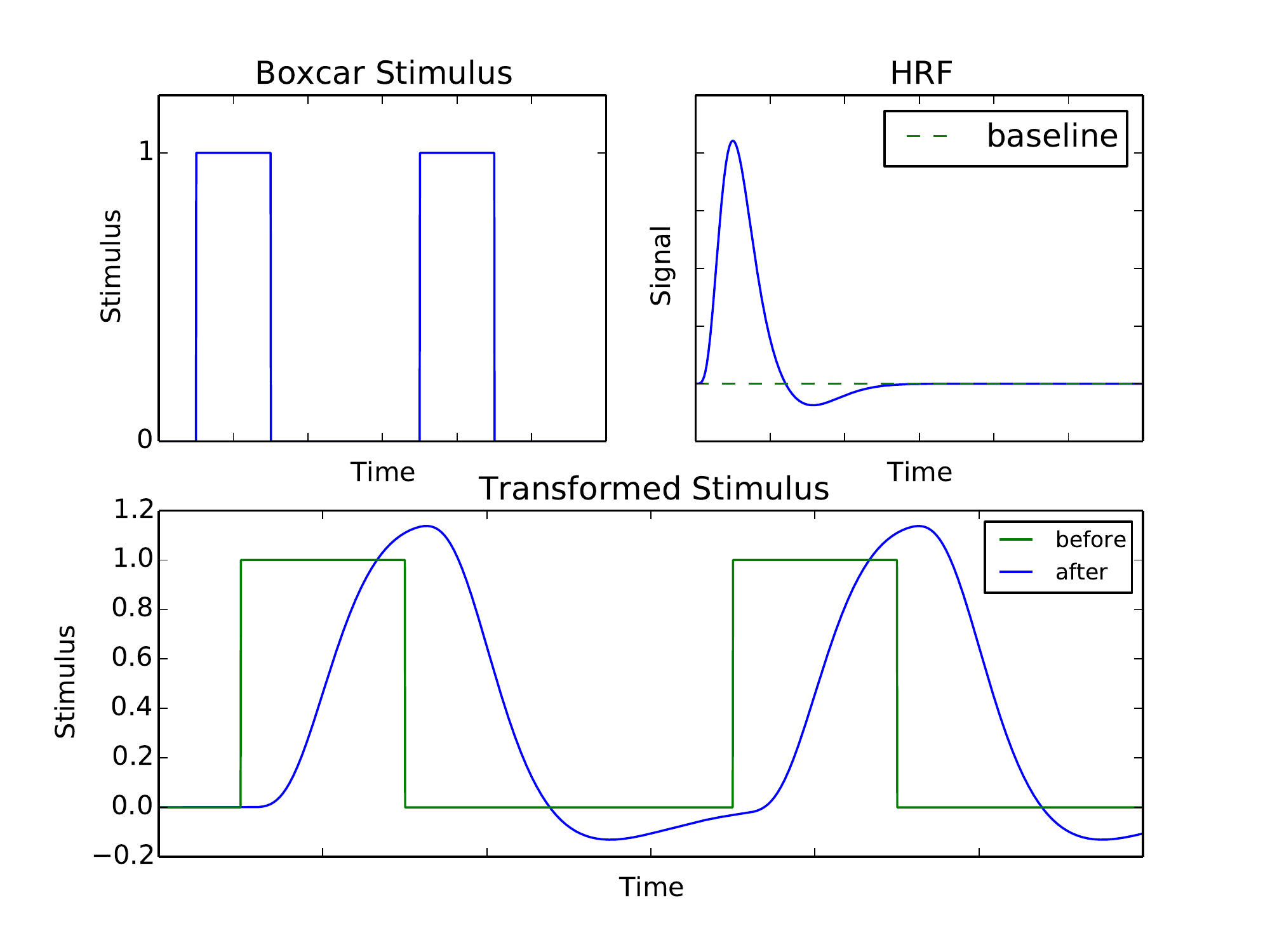}
\caption{The transformed stimulus is obtained by convolving the original 0-1 'boxcar' stimulus and the HRF.}
\label{fig:stimulus}
\end{figure}

For voxel $v = 1, \dots, N$, let $\left\{y_{v, i}; i = 1, \dots, t \right\}$ be the BOLD image intensities at $t$ time points. Although other alternatives are possible, a conventional voxelwise regression analysis assumes a linear model with a balance between model complexity and computational feasibility \citep{smit:fahr:2007, fris:ashb:frit:poli:heat:frac:1995},
\[
y_{v, i} = z_i^{T} a_v + x_{v, i} \beta_{v} + \epsilon_{v, i}.
\]
Linear combination $z_i^{T} a_v$ is the baseline trend to remove stimulus-independent effects. $\beta_{v}$ is the activation amplitude and $x_{v, i}$ is the transformed stimulus (see Figure~\ref{fig:stimulus}). In many experiments, the external stimulus $\left\{ s_i; i = 1, \dots, t \right\}$ alternates activation/inactivation in a 0-1 'boxcar' pattern. However, instead of proceeding in a 0-1 'boxcar' function, the brain produces a fairly fixed, stereotyped blood flow response with delay $d_v$ every time a stimulus hits it, where $d_v$ is estimated in a preprocessing step. The so-called hemodynamic response function (HRF) is used to characterizes this process. There are several formulations of HRF \citep[see e.g.][]{fris:flet:jose:holm:rugg:turn:1998, goss:auer:fhar:2001, glov:1999}. 

One approach is to use a canonical HRF consisting of a difference of two gamma functions \citep{lind:meng:atla:wage:2009},
\[
h(t) = A\left( {t^{\alpha_1-1} \beta_1^{\alpha_1} e^{-\beta_1t} \over \Gamma(\alpha_1) }-c{t^{\alpha_2-1} \beta_2^{\alpha_2} e^{-\beta_2 t} \over \Gamma(\alpha_2) } \right),
\]
where $\alpha_1 = 6$, $\alpha_2 = 16$, $\beta_1 = \beta_2 = 1$ and $c = 1/6$. The only unknown parameter, i.e. the amplitude $A$, is estimated in a preprocessing step. We can transform the orignal 'boxcar' stimulus by a convolution with the HRF,
\[
x_{v, i} = \sum_{k = 0}^{i - d_v} h(k) s_{i-d_v-k}.
\]
The measurement error is denoted by $\epsilon_{v, i}$. Appropriate distributional assumptions about $\epsilon_{v, i}$ can be made to incorporate temporal correlation and specific priors can be chosen to reflect spatial dependence. 


In this article, we apply the spatial Bayesian variable selection models for single subject proposed by \cite{lee:jone:caff:bass:2011}. Here we summarize their model formulation and estimation process. An interested reader is directed to their paper for more details. 

Denote $\pmb{y}_v = \left( y_{v, 1}, \dots, y_{v, t} \right)^{T}$ as the BOLD image intensity at time $i = 1, \dots, t$ for voxel $v = 1, \dots, N$. Let $X_v$ be a $t \times p$ design matrix of transformed stimulus and $\pmb{\beta}_v = \left( \beta_{v, 1}, \dots, \beta_{v, p} \right)^{T}$ be a vector of $p$ regression coefficients for each voxel. We formulate a linear regression mode,
\begin{equation} 
\label{eq:lm}
\pmb{y}_v = X_v \pmb{\beta}_v + \pmb{\epsilon}_v, \hspace{5mm} \pmb{\epsilon}_v \sim N_t \left( \pmb{0},  \sigma_v^2 \Lambda_v \right).
\end{equation}

Notice that the detection of voxel activation is equivalent to the identification of nonzero $\pmb{\beta}_v$s. To this end, we introduce 0/1 binary indicators $\pmb{\gamma}_v = \left( \gamma_{v, 1}, \dots, \gamma_{v, p} \right)$, $v = 1, \dots, N$, such that $\beta_{v, j} = 0$ if $\gamma_{v, j} = 0$ and $\beta_{v, j} \ne 0$ if $\gamma_{v, j} = 1$. The $\gamma_{v, j}$ is used to indicate whether the voxel $v$ is activated by input stimulus $j$. Given $\pmb{\gamma}_v$, let $\pmb{\beta}_v \left( \pmb{\gamma}_v \right)$ be the vector of nonzero regression coefficients and $X_v \left( \pmb{\gamma}_v \right)$ be the corresponding design matrix. Then, the model \eqref{eq:lm} can be rewritten as 
\begin{equation*}
\pmb{y}_v = X_v\left( \pmb{\gamma}_v \right) \pmb{\beta}_v \left( \pmb{\gamma}_v \right) + \pmb{\epsilon}_v.
\end{equation*}

Further, we assume the independence among $\sigma_v^2$ and set its prior $\pi \left( \sigma_v^2 \right) \propto {1 / \sigma_v^2}$. Zellner's $g$-prior on $\pmb{\beta}_v \left( \pmb{\gamma}_v \right) | \pmb{\gamma}_v$ is placed to undertake variable selection or model averaging. The parameter $g$ is adjusted to obtain similar results with those if BIC were used,
\begin{equation*} 
\pmb{\beta}_v \left(\pmb{\gamma}_v\right) |\pmb{y}_v, \sigma_v^2, \Lambda_v, \pmb{\gamma}_v \sim N\left(\hat{\pmb{\beta}}_v\left(\pmb{\gamma}_v\right), T_v\sigma_v^2\left[ X_v^T\left(\pmb{\gamma}_v\right) \Lambda_v^{-1} X_v\left(\pmb{\gamma}_v\right) \right]^{-1}  \right),
\end{equation*}
where
\begin{equation}
\hat{\pmb{\beta}}_v\left(\pmb{\gamma}_v\right) = \left[ X_v^T\left(\pmb{\gamma}_v\right) \Lambda_v^{-1} X_v\left(\pmb{\gamma}_v\right) \right]^{-1} X_v^T\left(\pmb{\gamma}_v\right) \Lambda_v^{-1} \pmb{y}_v.
\end{equation}
Define the corresponding sum of squares for posterior inference
\begin{equation*}
S\left( \rho_v, \pmb{\gamma}_v \right) = \left( \pmb{y}_v - X_v \left( \pmb{\gamma}_v \right) \hat{\pmb{\beta}}_v\left(\pmb{\gamma}_v\right) \right)^{T} \Lambda_v^{-1} \left( \pmb{y}_v - X_v \left( \pmb{\gamma}_v \right) \hat{\pmb{\beta}}_v\left(\pmb{\gamma}_v\right) \right).
\end{equation*}

We incorporate the temporal dependence between observations on a given voxel through the specification of the structure of $\Lambda_v$. The AR$(1)$ dependence, i.e. $\Lambda_v \left( i, j \right) = \rho_v^{|i-j|}$, is an effective compromise between inferential efficacy and computational efficiency. We specify a point mass prior for $\pmb{\rho} = \left( \rho_1, \dots, \rho_N \right)$ at a fixed point $\hat{\pmb{\rho}}$ using maximum likelihood methods.

We incorporate the spatial dependence, as well as the anatomical information, by using a binary MRF prior, i.e. Ising prior, on $\pmb{\gamma}_v$. Let $\pmb{\gamma}_{(j)} = \left( \gamma_{1, j}, \dots, \gamma_{N, j} \right)^{T}$ be the vector of indicators for regressor $j$ over all voxels. Then, let $w_{v, k}$ be pre-specified constants that weigh the interaction between voxels $v$ and $k$ and let $\nu_j$ be parameter to measure the strength of the interaction between voxels for regressor $j$. We denote $v \sim k$, if two voxels $v$ and $k$ are defined as neighbors by the user. In this article, we employ a widely used three-dimensional structure containing the six immediate neighbors: 1 above, 1 below and 4 adjacent. The weight $w_{v, k}$ is set to be the reciprocal of the Euclidean distance between voxel $v$ and $k$. Then, the spatial interaction is described as $\nu_j \sum_{v = 1}^{N} \sum_{v \sim k} w_{v, k} I \left( \gamma_{v, j} = \gamma_{k, j} \right)$, where $I (x)$ is the usual 0/1 indicator function. A linear "external field" $\sum_{v = 1}^{N} \alpha_{v, j} \gamma_{v, j}$ is specified to incorporate anatomical prior information, where $\alpha_{v, j}$ is chosen to reflect prior knowledge.

We consider the prior on $\pmb{\gamma}$ to be $\pi \left( \pmb{\gamma} | \pmb{\nu} \right) = \prod_{j = 1}^{p} \pi \left( \pmb{\gamma}_{(j)} | \nu_j \right)$, where
\begin{equation*}
 \pi \left( \pmb{\gamma}_{(j)} | \nu_j \right) \propto \exp \left\{ \sum_{v = 1}^{N} \alpha_{v, j} \gamma_{v, j} + \nu_j \sum_{v = 1}^{N} \sum_{v \sim k} w_{v, k} I \left( \gamma_{v, j} = \gamma_{k, j} \right) \right\}.
\end{equation*}

The remaining prior to be addressed is the distribution of $\pmb{\nu} = \left( \nu_1, \dots, \nu_p \right)$. A uniform prior is placed $\pi \left( \pmb{\nu} \right) \propto \prod_{j = 1}^{p} I \left( 0 < \nu_j < \nu_{max} \right)$, where \cite{moll:2003} suggests to use $\nu_{max} \le 2.0$.

The posterior density is characterized by
\begin{equation*}
q\left( \pmb{\beta} \left( \pmb{\gamma} \right), \pmb{\gamma}, \pmb{\rho}, \pmb{\nu}, \pmb{\sigma}^2 | y \right) \propto p\left( y |  \pmb{\beta} \left( \pmb{\gamma} \right), \pmb{\gamma}, \pmb{\sigma}^2, \Lambda \right) \quad \times  \pi\left( \pmb{\beta} \left( \pmb{\gamma} \right) | y, \pmb{\sigma}^2, \Lambda, \pmb{\gamma} \right) \pi\left( \pmb{\gamma} | \pmb{\nu} \right) \pi\left( \pmb{\rho} \right) \pi \left( \pmb{\sigma}^2 \right) \pi \left( \pmb{\nu} \right) .
\end{equation*}

We follow the two-step component-wise Metropolis-hastings algorithm designed by \cite{lee:jone:caff:bass:2011} to update $\pmb{\gamma}$ and $\pmb{\nu}$. Particularly, we are interested in estimating the posterior mean of $\pmb{\theta} = \{ \pmb{\gamma}, \pmb{\nu} \}$.

\begin{figure}[thb]
\centering
\includegraphics[width=0.7\textwidth]{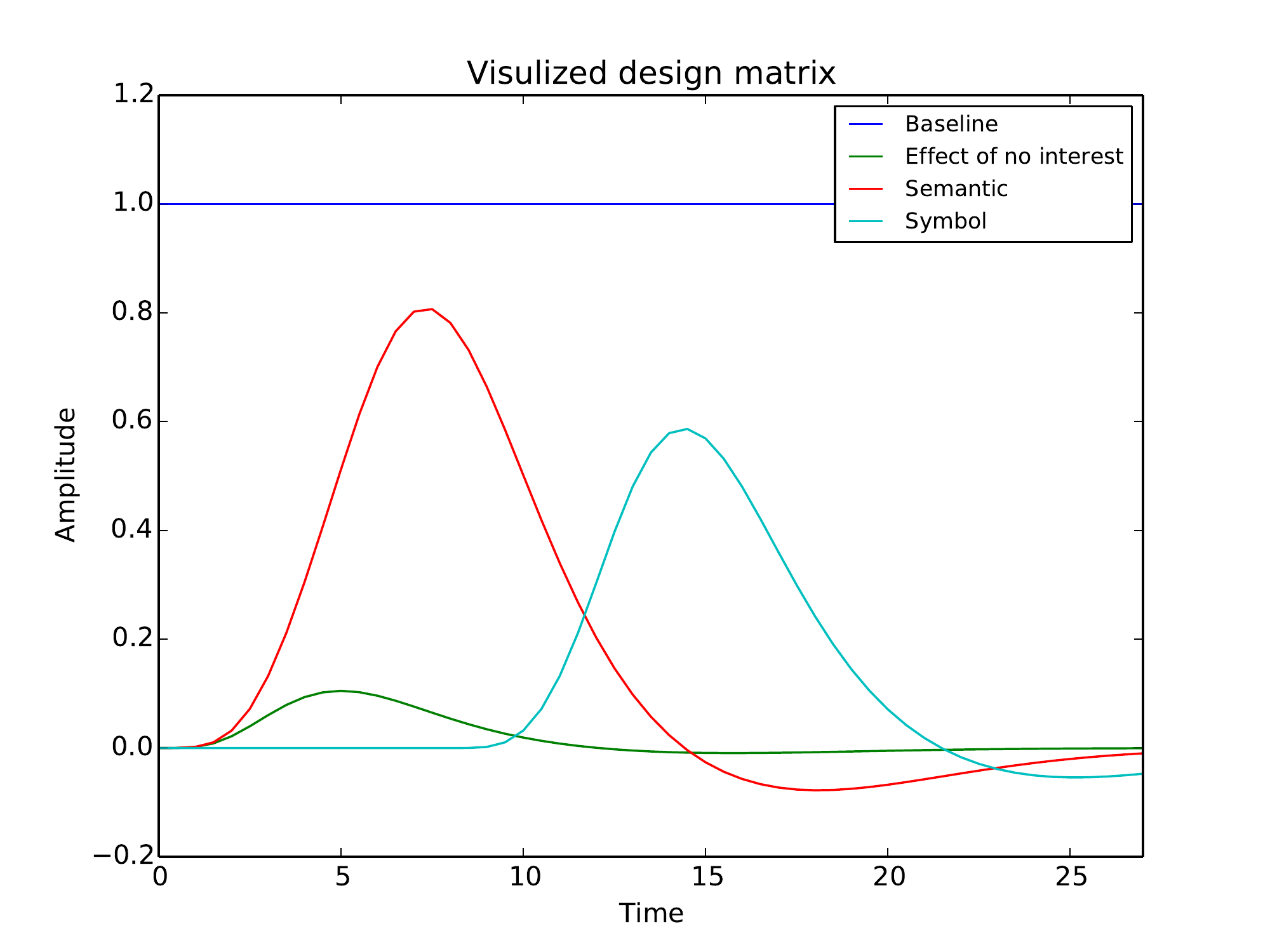}
\footnotesize \caption{The visualization of the design matrix for the experimental dataset.}
\label{fig:design}
\end{figure}

We rewrite the linear model \eqref{eq:lm} as
\[
\pmb{y}_v = \alpha_0 \pmb{z}_0 + \alpha_1 \pmb{z}_1 + \beta_1 \pmb{x}_1 + \beta_2 \pmb{x}_2 + \pmb{\epsilon}_v,
\]
where $\alpha_i \pmb{z}_i$s are the baseline signal, $\beta_i$s are the activation amplitude corresponding to the two tasks "Semantic" and "Symbol", respectively, The binary indicator $\gamma_v = \left\{ 1, 1, \gamma_{v, 3}, \gamma_{v, 4} \right\}$ is used in the variable selection problem described previously. Notice that we assume all $\alpha_i$s nonzero and set $\nu_{max} = 1.0$ as in \cite{lee:jone:caff:bass:2011}. Figure~\ref{fig:design} visualize the design matrix for this linear model as we described previously.

We implemented the relative standard deviation FWSR in the simulation study. Provided the ultrahigh-dimensional nature of the posterior, memory issues forbid us to use the uBM for $T(\epsilon)$. Again, because of the high-dimensional nature of the problem, we set $\epsilon = 0.05$. We set $n^{*} =2^{14} =16,384$ and added 20 or 21 batches between checks. The nominal level of the coverage probabilities was set to $0.95$. 

\begin{table}[thb]
\begin{center}
\begin{tabular}{l|c}
Variable & $\nu_j$ \\
\hline
	Semantic & 0.500 (0.0005) \\
	Symbol & 0.499 (0.0005) \\
\hline
Length & 292,864 \\
\end{tabular}
\caption{Summary of $\hat{\nu}_j$s, based on $T(\epsilon = 0.05)$.}
\label{tab:inter}
\end{center}
\end{table}   

Table~\ref{tab:inter} summarizes the estimates of the strength of interaction (MCSEs are given in the parentheses) and total simulation effort. Notice that relatively small MCSEs indicate high estimation precision level.   

%

\begin{figure*}[thb]
    \subfloat[Task "Semantic"\label{subfig:s}]{%
      \includegraphics[width=0.5\textwidth]{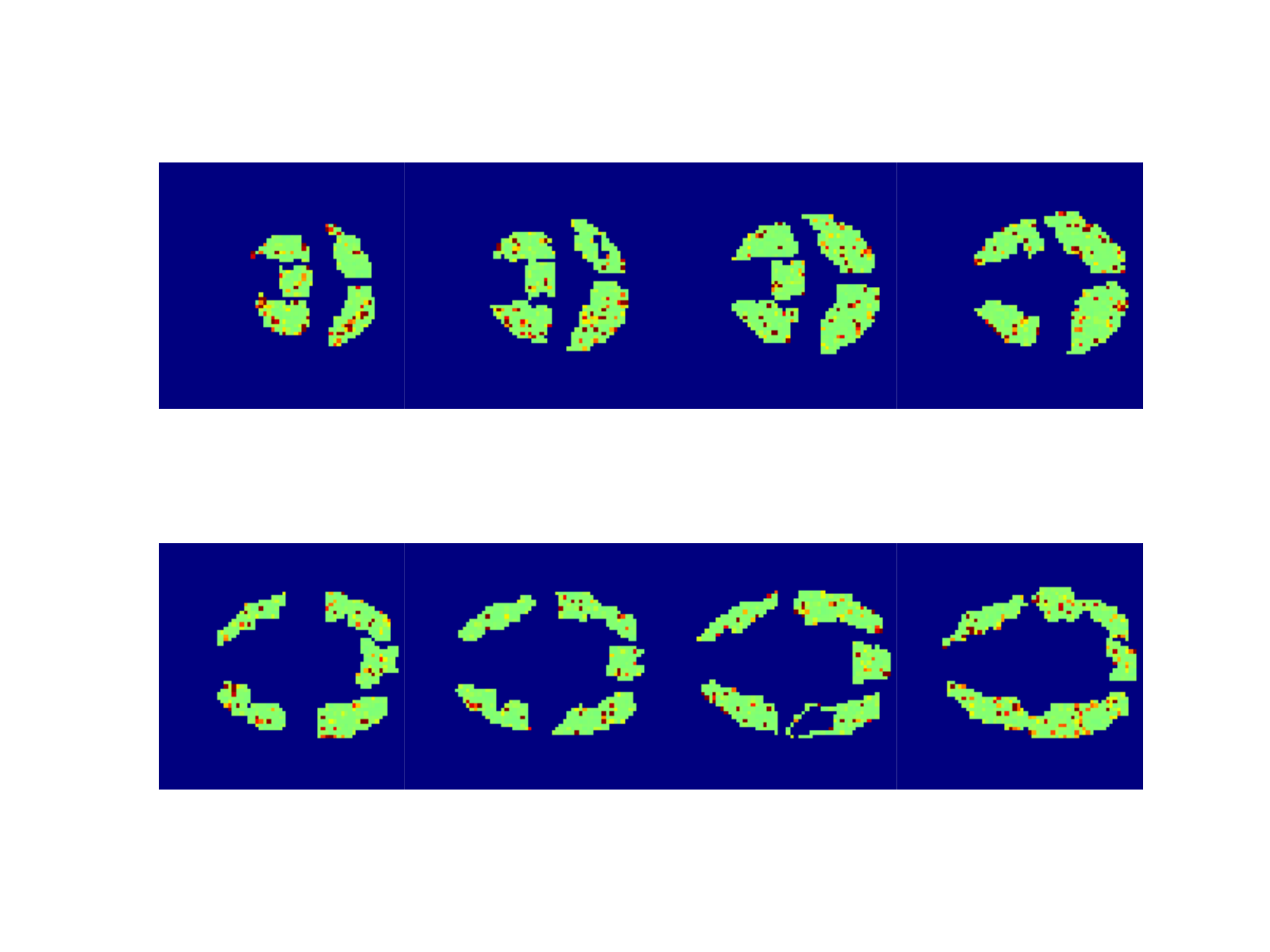}
    }
    \hfill
    \subfloat[Task "Symbol"\label{subfig:p}]{%
      \includegraphics[width=0.5\textwidth]{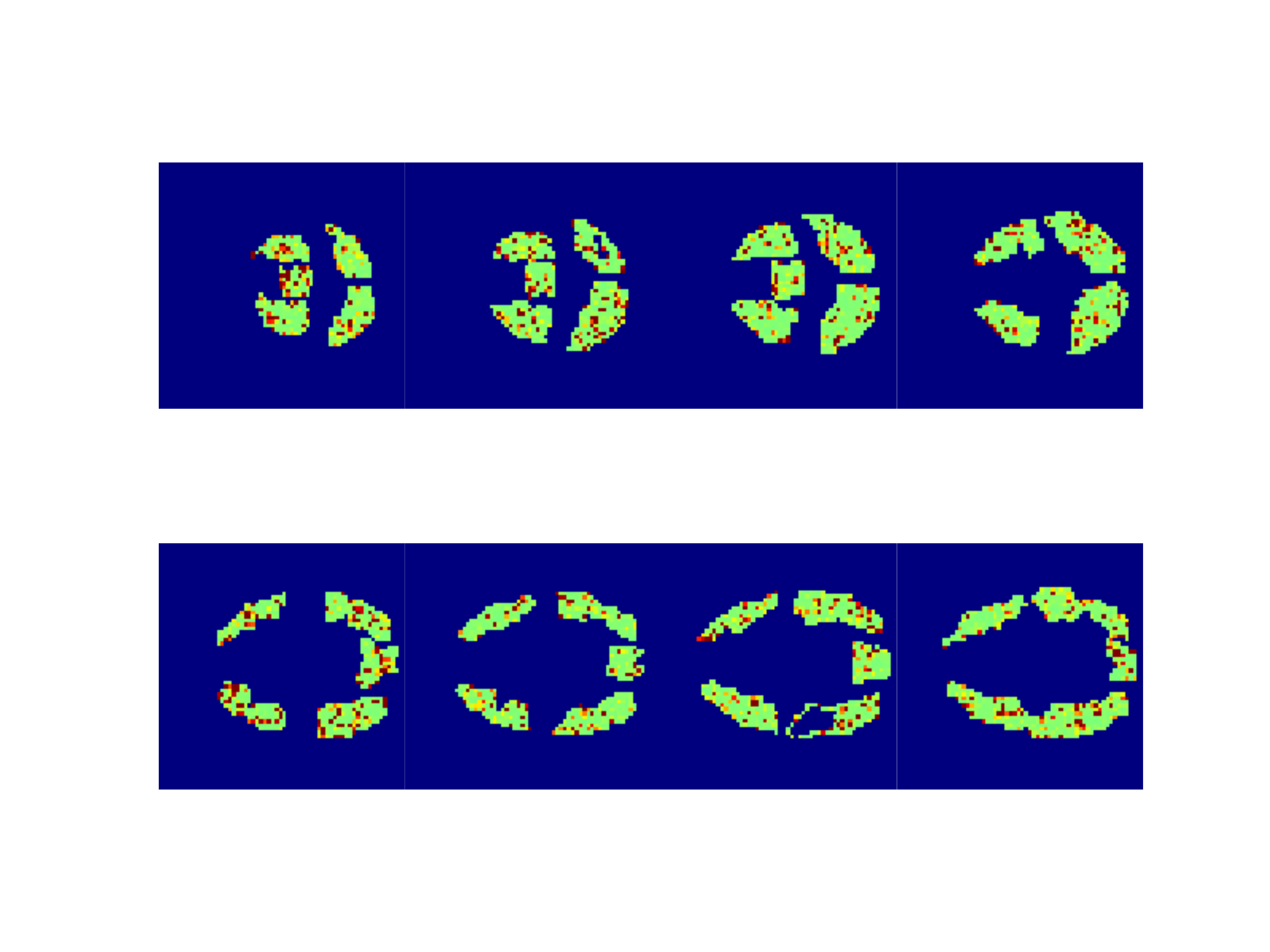}
    }
    \caption{The activation map for all eight slices when perform different tasks.}
    \label{fig:ps}
  \end{figure*}

Figure~\ref{subfig:s} and \ref{subfig:p} are the activation maps in all eight slices for two tasks "Semantic" and "Symbol", respectively. The red voxels are identified as activated. They gather into small clusters indicating the spatial correlation.

Again, we compare terminating simulations by $T(\epsilon)$ and through the GD. For the GD, we generated 15,000 iterations and used a $p$-value 0.05 to confirm the chain convergence for each voxel. Note that $p = 9398$ ($= 2 \times 4698 + 2$) versus the $15,000$ iterations in calculation. In such a short sample, we note a small proportion of the voxels stays active or inactive throughout the sample, i.e. a sequence of constant 0 or 1. Although we did not take this scenario into the comparison, it may suggest premature termination in some senses. 

\begin{figure}[thb]
\centering
\includegraphics[width=0.7\textwidth]{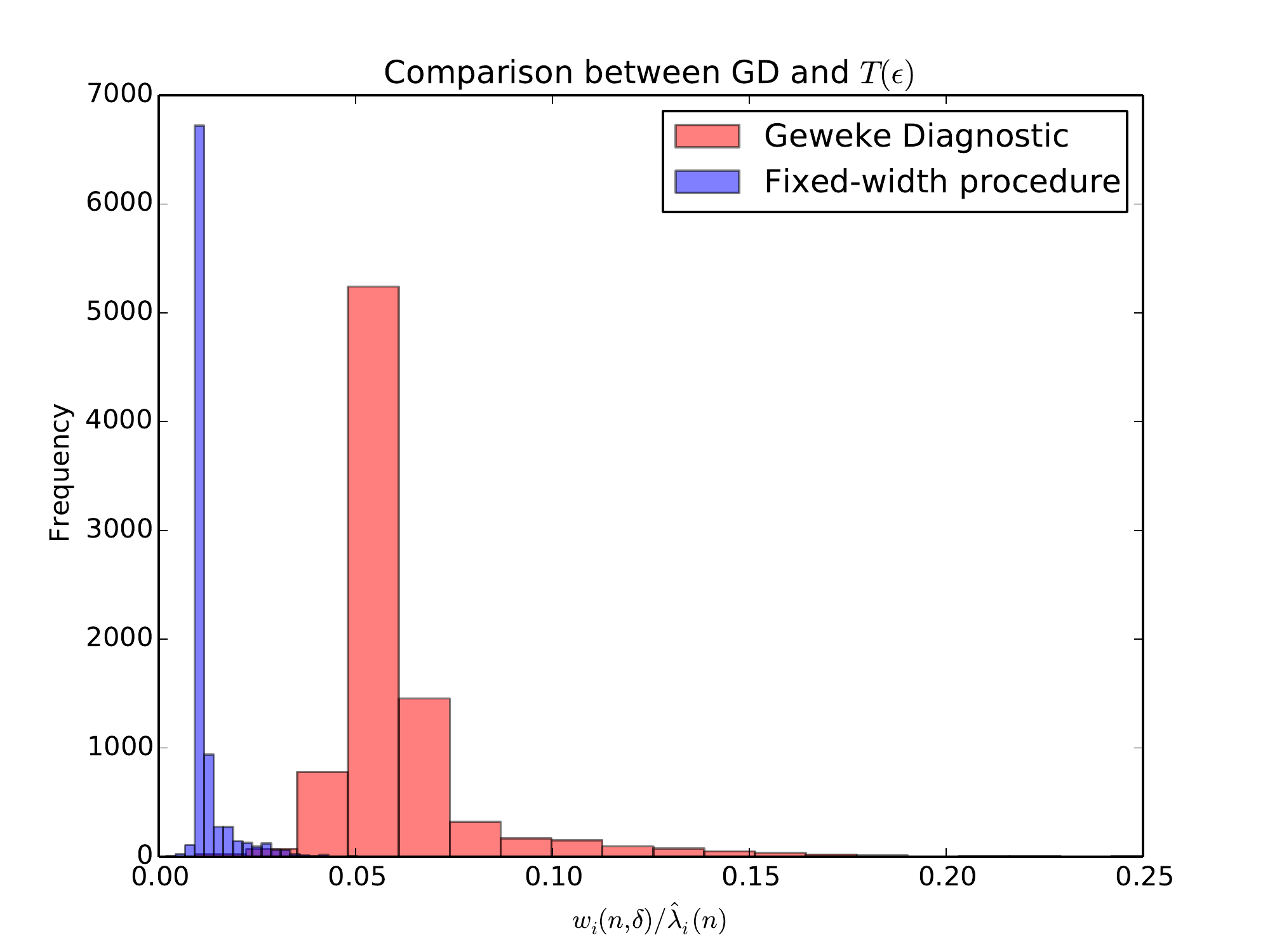}
\footnotesize \caption{The visualization of the comparison between GD and $T(\epsilon)$ in terms of the ratios $w_{i} (n, \delta) / \hat{\lambda}_{i} (n)$.}
\label{fig:compare}
\end{figure}

As before, we compare the two approaches in terms of the ratios $w_{i} (n, \delta) / \hat{\lambda}_{i} (n) $. In this example, we attempted to accurately estimate the true posterior standard deviation, $\lambda_{i}$, from a separate simulation with $10^7$ iterations. Figure~\ref{fig:compare} visualizes the comparison. It is clear that the ratios from $T(\epsilon)$ are significantly smaller and more concentrated than the ones from GD. However, $T(\epsilon)$ did require a greater simulation effort (292,864) than did GD (15,000), which might be a concern given the expensive computational cost to update the chain. However, one can adjust $\epsilon$ to reach a compromise between the accuracy of estimates and the required simulation effort.

\section{Discussion} \label{sec:disc}

This paper considers the relative FWSRs in the context of truly high-dimensional MCMC simulations. In our viewpoint, a practical stopping rule should achieve three properties: (1) it is easy to implement in an automated fashion with a few tuning parameters; (2) it attains confidence in resulting estimates; and (3) it is applicable in both low- and high-dimensional settings. With such properties, practitioners can then apply the stopping rule on a routine basis. 

We advocate use of the modified relative standard deviation FWSR since it meets all the properties and is especially applicable in high-dimensional Bayesian settings without prior knowledge of the magnitude of the target parameters. It is controlled by one tuning parameter $\epsilon$ that measures the accuracy of the estimates. Simply put, the estimates are approximately $\epsilon^{-1}$ more accurate than their posterior standard deviation. Another way to understand the $\epsilon$ is through the alternative ESS calculation. \cite{fleg:gong:2013} suggest to set the tuning parameter $\epsilon = 0.02$ which leads to an ESS of 38,416 (with $\delta=0.05$). They show that such a $\epsilon$ results in confidence intervals with desired coverage probabilities. However, the value of $\epsilon$ significantly affect the total simulation effort. For instance, in the described stimulation study, $\epsilon = 0.05$ results in 292,864 iterations, while $\epsilon = 0.02$ results in 1,419,264 iterations. Thus, there should be a balance between the accuracy of estimation and the cost of simulation, which depends on the estimation problem of interest. For the estimation of single posterior mean, we find $\epsilon = 0.05$ and $\delta=0.05$, or equivalently ESS=6,147, works well.

Recall the proposed sampling plan aBM summarizes information along the simulation. On one hand, it eliminates the requirement of storing the entire chain, solves the memory issues and reduces the computational time. On the other hand, it limits the use of such method to more general estimation problems, such as quantile estimation \citep{fleg:jone:neat:2012}. Certain tradeoffs are necessary in order to overcome obstacles arise from high-dimensionality.

One natural extension of the stopping rule is to consider simultaneous multivariate estimation. \cite{fleg:gong:2013} apply a Bonferonni approach to adjust for multiplicity. However, the standard Bonferonni approach will not work for a large dimension $p$, since the individual confidence interval needs to be set to a nominal level of $0.95^{1/p}$. A sophisticated method is required to adjust for multiplicity. One direction of future research is to control the volume of a desired confidence region rather than the width of multiple confidence intervals separately.

Finally, how well a chain mixes and explores the state space is by all means an vital component in MCMC simulations. Especially in high dimensional settings, it is still a challenging problem to date. In the fMRI application, we found a small proportion of the 0-1 type chains stay at a fixed point for a large number of iterations suggesting poor mixing. Interested readers can refer to \cite{broo:gelm:jone:meng:2010} for more information.

\section*{Acknowledgements}
The authors thank Brian Caffo and Galin Jones for helpful conversations about this paper.  The first author is grateful to the Section on Bayesian Statistical Science of the ASA, who recognized an earlier version of this paper as a winner of their anual student paper competition.  The second author's work is partially supported by NSF grant DMS-13-08270.

\setstretch{1}
\bibliographystyle{apalike}
\bibliography{ref}

\end{document}